\documentclass[aps,pra,reprint,amsmath,amssymb]{revtex4-2}
\usepackage{graphicx}
\usepackage{hyperref}
\usepackage{natbib}
\usepackage{color}
\hypersetup{colorlinks,allcolors=blue,bookmarksnumbered}
\usepackage{xcolor}
\colorlet{RevColor}{red}

\begin{document}
\title{Phase transition, phase separation and mode softening of a two-component Bose-Einstein condensate in an optical cavity}
\author{Jia-Ying Lin}\altaffiliation{These authors contributed equally to this work.}
\author{Wei Qin}\altaffiliation{These authors contributed equally to this work.}
\author{Renyuan Liao}
\email{ryliao@fjnu.edu.cn}
\affiliation{College of Physics and Energy, Fujian Normal University, Fujian Provincial Key Laboratory of Quantum Manipulation and New Energy Materials, Fuzhou 350117, China}
\affiliation{Fujian Provincial Engineering Technology Research Center of Solar Energy Conversion and Energy Storage, Fuzhou 350117, China}
\date{\today}

\begin{abstract}
We investigate the superradiant phase transition in a two-component Bose-Einstein condensate with distinct atomic detunings, confined in an optical cavity and driven by a transverse pump laser. By combining perturbation theory and numerical simulations, we demonstrate that the phase transition is dominated by the red-detuned component, resulting in a phase diagram completely different from that of a single-component case under blue-detuned condition. The system exhibits spontaneous phase separation between the two components, manifested as alternating stripe patterns in the normal phase and distinct Bragg gratings in the superradiant phase. Furthermore, the Bogoliubov excitation spectrum reveals roton-type mode softening, indicating that the phase transition also corresponds to the superfluid-to-lattice supersolid transition. Our findings provide insights into the interplay between atomic detunings and collective quantum many-body phenomena, offering potential applications in quantum simulation and optical switching technologies.
\end{abstract}

\maketitle

\section{Introduction}
The preparation of ultracold atoms within optical cavities has opened new frontiers for exploring collective many-body phenomena~\cite{Bloch2008,Ritsch2013,Farokh2021}. The atom-light interaction in such systems can generate highly nonlocal nonlinearity, leading to various novel phenomena~\cite{Black2003,Summy2016,Landig2016,Dalafi2017,Ghasemian2019,LiZhengChun2019,Sauerwein2023,Helson2023,Defenu2023}.
One of the most prominent examples is the realization of Dicke model and the observation of self-organized phase~\cite{Baumann2010,ZhangYuanWei2013,ChenYu2014,ChenYu2015,Leonard2017,Zupancic2019,LiXiangLiang2021,NieXiaotian2023,ChenShi2024,QinWei2025}. It is experimentally manifested as photons being scattered into the cavity field accompanied by periodic arrangement of particles and the core mechanism for this phenomenon can be explained in terms of superradiance.
Both theoretical and experimental studies have demonstrated that the superradiant phase transition in single-component Bose-Einstein condensate (BEC) depends on the transverse pump lattice potential experienced by atoms~\cite{Baumann2010,Zupancic2019,QinWei2024}: For red atomic detuning, the phase transition occurs when the potential gain is sufficient to overcome kinetic energy loss, and the collective excitation mode corresponding to the coupling momentum softens at the critical point~\cite{Nagy2008,Mottl2012,Julian2017}; For blue atomic detuning, the antisymmetric coupling to the $P$-band of the pump lattice induces self-organization. But the system leaves the superradiant phase at high pump lattice depth due to vanishing overlap with the $P$-band~\cite{Zupancic2019}. In addition, this self-consistent ordering of atoms and light has become a paradigmatic system for exploring driven-dissipative quantum phases~\cite{Nagy2009,Vaidya2018,Mivehvar2018,Nishant2019,LinRui2020,GaoPan2023,Tolle2025,ZhangZhao2025,Ericsong2025}.

The inclusion of internal atomic spin degrees of freedom in BEC-cavity systems has been widely studied~\cite{Safaei2013,Mivehvar2017,Chiacchio2019,Ostermann2021,Carl2023,Mivehvar2024,Chelpanova2024}. Recently, a two-component BEC system implemented via this approach demonstrates superradiant phenomena with density and spin self-organization, achieved via the vectorial atom-light interaction between different internal states of a driven BEC and the vacuum mode of a cavity~\cite{Landini2018,Kroeze2018,Ferri2021}.  A limitation of such studies on superradiant phase transitions is that the energy-level selection typically results in identical atomic detuning for all components, thereby overlooking the potential influence of distinct detunings.
%However, such studies on superradiant phase transition typically neglect the effects of different atomic detunings between components due to the assumption that these detunings are identical.
An alternative approach for constructing a two-component BEC involves using different atomic species, which may induce novel quantum phases. The underlying reason is that the parameters such as inter-component energy level differences, mass differences, interaction strengths and so on can be crucial in breaking some of the existing symmetries of the system~\cite{Ali2022,ZhengYongGuang2025}. However, theoretical studies of phase transition and phase diagrams for superradiance in such systems remain scarce. Here, we are trying to fill in this gap by investigating such a two-component system with different atomic detunings using two BECs at distinct energy levels. Specifically, we shall carry out a comprehensive theoretical calculations on the ground-state structures, critical behaviors  and  associated quantum phase transitions, when the two-component BEC is placed in a single-mode optical cavity and driven by a transverse pumping laser.
%Due to growing theoretical interest and immediate experimental relevance, we shall address this issue by inverstigating such a two-component system with different atomic detunings using two BECs at distict energy levels, and shall theoretically describe the superradiant phase transition when the two-component BEC is placed in a single-mode optical cavity and driven by a transverse pumping laser.

This paper is organized as follows. First, we will lay out the model of our problem, present the theoretical formulation starting from the model Hamiltonian, and derive analytical expressions associated with phase transitions of the system by the perturbation theory. Second, by combining analytical and numerical results, we will map out the phase diagrams and analyze which is the dominant factor in driving the phase transition. Third, we will monitor the evolution of the order parameters by varying the tuning parameter, demonstrating the existence of a natural phase separation between the two components. Fourth, by examining the Bogoliubov excitation spectrum, we show that the softening of the roton-type mode indicates a phase transition from a superfluid to a lattice supersolid. Finally, we will conclude with a summary and outlook.

\section{MODEL and theoretical formulation}
\begin{figure}[t]
    \includegraphics[width=1\linewidth]{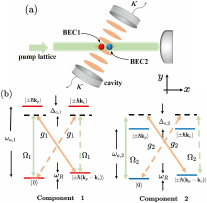}
    \caption{(a) Schematic illustration of the two-component BEC trapped inside a high-finesse optical cavity (the cavity field is drew by orange stripes) and driven by a transverse pump (the pump field is drew by green long-line), where red and blue balls represent single-component BECs with red ($\Delta_{a,1}<0$) and blue detunings ($\Delta_{a,2}>0$), respectively. The cavity field decays at a rate of $\kappa$. Here the angle between the cavity beam and pump beam is $60^\circ$, and both the pump and cavity fields are polarized orthogonal to the $x$-$y$ plane. (b) The scattering paths of momentum state can be visualized for two components. Light scattering between the pump field and the cavity mode induces Raman couplings between the zero momentum state $|\mathbf{p}\rangle=|0\rangle$ and the excited state $|\pm\hbar(\mathbf{k}_p-\mathbf{k}_c)\rangle$ at energy $E_R=\hbar\omega_R$. $\Omega_j$ and $g_j$ represent respectively the Rabi frequencies of pump laser and cavity mode, for components $j=1,2$. $\omega_{a,j}$ is the transition frequency of the two-level atoms, $\Delta_{a,j}$ is the atom-pump detuning, and $\Delta_c$ is the cavity-pump detuning.  The colors in the energy level correspond to the colors in the schematic diagram. It is not assumed here that the ground states corresponding to $|0\rangle$ of components 1 and 2 are at the same level.}
    \label{fig1}
\end{figure}

We consider that a two-component BEC is trapped inside a high-finesse optical cavity and illuminated by a transverse pump at an angle of $60^\circ$ to the cavity axis, as sketched in Fig.~\ref{fig1}\textcolor{blue}{(a)}. The cavity mode frequency is $\omega_c$, the pump frequency is $\omega_p$, and the transition frequency of the two-level atoms is $\omega_{a,j}$ with components $j=1$ and $2$. Here, The cavity-pump detuning and the atom-pump detuning are represented as $\Delta_c=\omega_p-\omega_c$ and $\Delta_{a,j}=\omega_p-\omega_{a,j}$, respectively. We assume that the pump beam is detuned far from the atomic transition $\omega_{a,j}$, however the component $1$ experiences an attractive pump potential $(\Delta_{a,1}<0)$ and the component $2$ experiences a repulsive pump potential $(\Delta_{a,2}>0)$. The polarization direction of the pump field is parallel to that of the cavity mode, with both oriented  orthogonal to the $x$-$y$ plane. Thus, the contribution of vector light-atom interaction to the superradiative phase transition vanishes~\cite{Zupancic2019, New_Morales2019}. So that we can adiabatically eliminate the electronic excited states of the two components, and obtain an effective Hamiltonian in a frame rotating at the pump laser frequency:
\begin{equation}
\hat{\mathcal{H}}=\sum_{j=1,2}\int\hat{\Psi}_j^\dagger \hat{H}_{0,j}\hat{\Psi}_jd\mathbf{r}-\hbar\Delta_c\hat{a}^\dagger\hat{a},
\label{eq:effective_Hamiltonian_in_a_frame_rotating}
\end{equation}
\begin{equation}
\hat{H}_{0,j}=\hat{H}_{at,j}+\mathrm{sgn}(\Delta_{a,j})\hbar\big[U_j(\mathbf{r})\hat{a}^\dagger\hat{a}+\eta_j(\mathbf{r})(\hat{a}^\dagger+\hat{a})\big],
\end{equation}
\begin{equation}
\hat{H}_{at,j}=\frac{\mathbf{p}^2}{2m_j}+\mathrm{sgn}(\Delta_{a,j})\hbar V_j(\mathbf{r}),
\end{equation}
where $\hat{\Psi}^\dagger_j$ $(\hat{\Psi}_j)$ is the atomic creation (annihilation) operator of component $j$ and $\hat{a}^\dagger$ $(\hat{a})$ is the photon creation (annihilation) operator for the cavity mode. The optical potentials  $V_j(\mathbf{r})$ and $U_j(\mathbf{r})$ are generated by the pump beam and the cavity field, respectively. We have $V_j(\mathbf{r})=V_{0,j}\cos^2(\mathbf{k}_p\cdot\mathbf{r})$,  $U_j(\mathbf{r})=U_{0,j}\cos^2(\mathbf{k}_c\cdot\mathbf{r})$ with $V_{0,j}=\Omega^2_j/|\Delta_{a,j}|$ and $U_{0,j}=g^2_j/|\Delta_{a,j}|$. The interference between the pump beam and the cavity field gives rise to $\eta_j(\mathbf{r})=\eta_{0,j}\cos(\mathbf{k}_p\cdot\mathbf{r})\cos(\mathbf{k}_c\cdot\mathbf{r})$ with $\eta_{0,j}=g_j\Omega_j/|\Delta_{a,j}|$. Here $\Omega_j$ is the strength of the pump beam, $g_j$ is the single-photon Rabi frequency of the cavity mode, $\mathbf{k}_p=k_0\hat{x}$ is the wave vector of the pump beam and $\mathbf{k}_c=k_0\cos(60^\circ)\hat{x}+k_0\sin(60^\circ)\hat{y}$ is the wave vector of the cavity light. We define the recoil energy $E_R=\hbar^2k_0^2/2m$ as the energy unit and $\mathrm{sgn}(\Delta_{a,j})$ gives the sign of detuning $\Delta_{a,j}$. For simplicity, we set $\hbar=1$, $m_j=m$ from now on. Superradiance can be driven by increasing the pump beam potential $V_{0,j}$, which simultaneously increases $\eta_{0,j}$ through the relation $\eta_{0,j}=\sqrt{V_jU_j}$.

The weak transmission of the cavity mirror leads to a small photon decay rate $\kappa$ for the cavity mode \cite{Baumann2010, ChenYu2014}. We implement the mean-field theory by replacing $\hat{a}$ with $\alpha=\langle \hat{a} \rangle$, and the dynamics of average photon number is derived from the Heisenberg equation of motion for the photon field operator:
\begin{equation}
i\frac{\partial \alpha}{\partial t}=-(\tilde{\Delta}_c+i\kappa)\alpha+\Theta,
\label{eq:photon_move_eq}
\end{equation}
with the effective cavity detuning $\tilde{\Delta}_c=\Delta_c-\langle\sum_i\mathrm{sgn}(\Delta_{a,j})\int U_j(\mathbf{r})\hat{N}_j\mathrm{d}\mathbf{r}\rangle$ and the density order parameter $\Theta=\langle\sum_j\mathrm{sgn}(\Delta_{a,j})\int \eta_j(\mathbf{r})\hat{N}_j\mathrm{d}\mathbf{r}\rangle$ with $\hat{N}_j=\hat{\Psi}^\dagger_j\hat{\Psi}_j$. In the absence of the cavity field, we can define an eigenstate of the system as a tensor product $|\psi^{n,n'}(\mathbf{k,k'})\rangle=|\Psi_1^n(\mathbf{k})\rangle \otimes |\Psi_{2}^{n'}(\mathbf{k}')\rangle$ where $\Psi_j^n(\mathbf{k})$ denotes the $n$th band eigenstate of the single-particle Hamiltonian $\hat{H}_{at,j}$. At the zero-temperature limit, all atoms for each components occupy the lowest-energy Bloch state $|\Psi_j^1(0)\rangle$ in the absence of intracavity photons. We seek a steady state in which $\partial \alpha/\partial t=0$ and obtain
\begin{equation}
\alpha=\frac{\Theta }{\tilde{\Delta}_c+i\kappa}.
\label{eq:stead_state_alpha}
\end{equation}

We can derive a Landau-type theory for the phase transition \cite{Morales2018}, and the phase boundary between the normal phase and the superradiant phase can be determined by the perturbation theory \cite{Zupancic2019, QinWei2024}. Up to quadratic order in $|\alpha|$, the ground-state energy $\varepsilon=\langle \hat{\mathcal{H}} \rangle $ is written as
\begin{equation}
\varepsilon =-4\mathcal{F}(Re\alpha)^2-\tilde{\Delta}_c|\alpha|^2,
\label{eq:perturbation_energy}
\end{equation}
with the susceptibility $\mathcal{F}$ given by
\begin{equation}
\mathcal{F}=\sum_{n,n',\mathbf{k},\mathbf{k'}}\frac{|\langle\psi^{n,n'}(\mathbf{k,k'})|\sum_{j}\mathrm{sgn}(\Delta_{a,j})\eta_j(\mathbf{r})\hat{N}_j|\psi^{1,1'}(\mathbf{0,0}) \rangle|^2}{E^{n,n'}(\mathbf{k},\mathbf{k}')-E^{1,1}(0,0)},
\label{eq:susceptibility_function}
\end{equation}
where $E^{n,n'}(\mathbf{k},\mathbf{k}')=E^{n}_{1}(\mathbf{k})+E^{n'}_{2}(\mathbf{k}')$ with $E^1_{j}(0)$ and $E^{n}_{j}(\mathbf{k})$ being the eigenvalues associated with $|\Psi^{1}_{j}(0) \rangle$ and $|\Psi^{n}_{j}(\mathbf{k})\rangle$, respectively. Substituting Eq.~(\ref{eq:stead_state_alpha}) into Eq.~(\ref{eq:perturbation_energy}), one finds that the ground-state energy can be expressed in terms of the order parameter $\Theta$ as
\begin{equation}
\varepsilon=-\frac{\tilde{\Delta}_c}{\tilde{\Delta}_c^2+\kappa^2}(1+\frac{4\mathcal{F}\tilde{\Delta}_c}{\tilde{\Delta}_c^2+\kappa^2})\Theta^2
\label{eq:energy_expressed_Theta}.
\end{equation}

\begin{figure}[t]
    \includegraphics[width=0.95\linewidth]{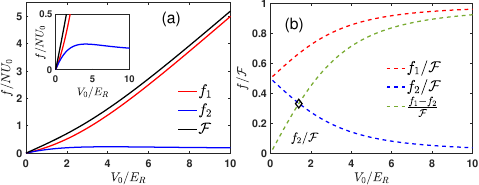}
    \caption{(a) Numerical results for individual components $f_j$ and total susceptibility $\mathcal{F}$ as a function of $V_0/E_R$, where $\mathcal{F}=\sum_jf_j$. The inset shows the details for all curves.  The curves of $\mathcal{F}$ and $f_1$ are very similar, both increase monotonically with $V_0/E_R$. In contrast, $f_2$ reaches its maximum at $V_0\approx 4E_R$ and subsequently decreases slowly for $V_0>4E_R$. (b) The evolution of the ratios $f_1/\mathcal{F}$, $f_2/\mathcal{F}$ and $(f_1-f_2)/\mathcal{F}$ with respect to $V_0/E_R$. The rhombus indicates a relationship $f_2=f_1-f_2$ at $V_0\approx1.4E_R$. The red and blue lines correspond to the components of $\Delta_{a,1}<0$ and $\Delta_{a,2}>0$. $N$ represents the atomic number in the corresponding system.}
%The red and blue lines correspond to the cases of $\Delta_{a,1}<0$ and $\Delta_{a,2}>0$, respectively. The solid line is for momentum state $(\mathbf{k}_p-\mathbf{k}_c)$, whereas the dashed line is for the momentum state $-(\mathbf{k}_p-\mathbf{k}_c)$.}
    \label{fig2}
\end{figure}

Thus, the phase transition occurs when the sign of the coefficient of
$\Theta^2$ in Eq.~(\ref{eq:energy_expressed_Theta}) changes, yielding the established condition for onset of the superradiant phase
\begin{equation}
-\frac{4\mathcal{F}\tilde{\Delta}_c}{\tilde{\Delta}_c^2+\kappa^2}>1.
\label{eq:phase_condition}
\end{equation}

The key ingredient of Eq. (\ref{eq:phase_condition}) is the susceptibility $\mathcal{F}$ of the normal phase, which characterizes the tendency of inducing superradiance. The larger $\mathcal{F}$ is, the greater the critical magnitude of effective cavity detuning $|\tilde{\Delta}_c|$ is. In Fig.~\hyperref[fig2]{2(a)}, we present numerical results for individual components $f_j$ and total susceptibility $\mathcal{F}$ at zero temperature in weak lattice limit, where $\mathcal{F}=\sum_jf_j$. For simplicity, we further set $V_{0,j}=V_0$ and $U_{0,j}=U_0$, which leads to $\eta_{0,j}=\eta_0$. Here we choose the parameters from Ref.~\cite{Zupancic2019} for verification, where the atomic number of each component $N_j=2.7\times10^5$, the recoil energy $E_R=2\pi\times3.77 \mathrm{kHz}$, the cavity decay rate $\kappa=2\pi\times147 \mathrm{kHz}$ and $U_0=0.012 E_R$ are kept fixed in  the following discussion.

 The atoms acquire  strong density modulation in the directions $\mathbf{k}=\pm(\mathbf{k}_p-\mathbf{k}_c)$~\cite{Zupancic2019}, and Fig.~\hyperref[fig1]{1(b)} shows the momentum state couplings of two components by two-photon processes. The inset of Fig.~\hyperref[fig2]{2(a)} provides a detailed view of the individual components $f_1$, $f_2$, and the total response $\mathcal{F}$. We observe that $f_1$ grows monotonically with $V_0$, whereas $f_2$ exhibits a maximum at $V_0 \approx 4E_R$ beyond which it slowly declines. To quantify the relative contributions, we plot the ratios $f_1/\mathcal{F}$, $f_2/\mathcal{F}$, and $(f_1-f_2)/\mathcal{F}$ in Fig.~\hyperref[fig2]{2(b)}. The ratio $f_j/\mathcal{F}$ measures the fractional contribution of $f_j$ to the total $\mathcal{F}$, while $(f_1-f_2)/\mathcal{F}$ serves as a measure for the disparity between the two components. The close agreement between $\mathcal{F}$ and $f_1$ evident in Fig.~\hyperref[fig2]{2(a)} is explained by the data in Fig.~\hyperref[fig2]{2(b)}. The evolution of these ratios with $V_0$ reveals that the relative weight of $f_1$ increases significantly. A key insight is gained at $V_0 \approx 1.4E_R$ (the rhombus in Fig.~\hyperref[fig2]{2(b)}), where the relation $f_1 = 2f_2$ is satisfied. For $V_0 > 1.4E_R$, the contribution of $f_1$ to $\mathcal{F}$ is more than double that of $f_2$. Furthermore, the growth rate of $f_1$ substantially exceeds that of $f_2$, leading to the conclusion that the total susceptibility $\mathcal{F}$ is dominated by the component with $\Delta_{a,1} < 0$.

%Since the atoms acquire  strong density modulation in the directions $\mathbf{k}=\pm(\mathbf{k}_p-\mathbf{k}_c)$~\cite{Zupancic2019}, we present only the susceptibility $f_j$ for these two momentum states in Fig.~\ref{fig2}. Here the red and blue lines correspond to the cases of red atomic detuning ($\Delta_{a,1}<0$) and blue atomic detuning ($\Delta_{a,2}>0$), respectively, while the solid and dashed lines are numerical results of momentum states $(\mathbf{k}_p-\mathbf{k}_c)$ and $-(\mathbf{k}_p-\mathbf{k}_c)$, respectively. We find that different momentum states have different susceptibilities. The susceptibility $f_1$ grows as $V_0$ increases, yet this monotonicity disappears for the susceptibility $f_2$ when $V_0$ reaches a finite value. Furthermore, the growth rate of $f_1$ is much larger than that of $f_2$, indicating that the total susceptibility $\mathcal{F}$ is dominated by the contribution from the component with $\Delta_{a,1}<0$.

\section{Phase diagram and Phase separation}
\begin{figure}[t]
    \includegraphics[width=1\linewidth]{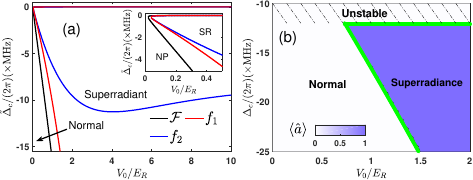}
    \caption{(a) Phase diagram as a function of the tuning parameter $V_0/E_R$ and the  effective cavity detuning $\tilde{\Delta}_c$, calculated from Eq~(\ref{eq:critical_effective_cavity_detuning}). Phase boundaries are shown for the binary mixture (black), a red-detuned single-component BEC (red), and a blue-detuned single-component BEC (blue). The inset shows a magnified view near $V_0\sim0.03E_R$. (b) Phase diagrams from the mean-field approach (green line) and the self-consistent numerical approach. The unstable region arises because the ground-state energy diverges with increasing order parameter $\langle \hat{a} \rangle$. The superradiant phase is defined by the condition $\langle \hat{a} \rangle \neq 0$.}
%Phase diagram spanned by the tuning parameter $V_0/E_R$ and cavity detuning $\Delta_c/(2\pi MHz)$. It contains three regions: the normal phase, the superradiant phase, and the unstable region. The green line is the phase boundary derived from Eq.~(\ref{eq:critical_effective_cavity_detuning}) and the actual phase boundary is determined by the self-consistent numerical approach. The dotted line corresponds to $|\Delta_c|<|\chi|$, where the system is unstable toward collapse. We define the superradiant phase region by condition $\langle \hat{a} \rangle \neq 0$.}
    \label{fig3}
\end{figure}

Solving the phase transition condition Eq.~(\ref{eq:phase_condition}) with respect to $\tilde{\Delta}_c$ yields the critical effective cavity detuning
\begin{equation}
\tilde{\Delta}_c^{\mathrm{crit}}=-2\mathcal{F}\pm\sqrt{4\mathcal{F}^2-\kappa^2},
\label{eq:critical_effective_cavity_detuning}
\end{equation}
Fig.~\hyperref[fig3]{3(a)} displays the mean-field phase diagram calculated from Eq. (\ref{eq:critical_effective_cavity_detuning}), showing the phase boundaries between the normal and superradiant phases. The results compare three distinct systems: a single-component BEC with red detuning ($\mathcal{F}=f_1$, red curve), a single-component BEC with blue detuning ($\mathcal{F}=f_2$, blue curve), and the binary mixture considered in this work ($\mathcal{F}=\sum_jf_j$, black curve).

We observe that the phase diagram for the binary mixture with distinct detunings closely resembles that of a single-component BEC under red detuning~\cite{Baumann2010}, apart from a slight horizontal shift to the left induced by $f_2$. This stands in stark contrast to the behavior of a blue-detuned single-component BEC, which typically reenters the normal phase at high lattice depths ($V_0>5-10E_R$)~\cite{Zupancic2019}. The system of two-component BEC with distinct detunings, however, sustains the superradiant phase even with a further increase in $V_0$. Furthermore, the critical lattice depth $V_0^{\mathrm{crit}}$ of the phase transition shifts to larger values with increasing  $|\tilde{\Delta}_c|$. The inset reveals the existence of a minimum threshold, here $V_0\sim 0.03E_R$,  below which superradiance does not occur. This critical value is determined by the condition $\mathcal{F}=\kappa/2$, which ensures that the argument of the square root in Eq.~(\ref{eq:critical_effective_cavity_detuning}) is positive.
%where we have additionally introduced the scattering matrix element $\chi=\sum_j\langle\Psi^{1}_{j}(0)|\mathrm{sgn}(\Delta_{a,j})U_j(\mathbf{r})|\Psi^{1}_{j}(0) \rangle$. Here we deal with only the situation where the cavity-pump detuning is negative $\Delta_c<0$.

The dynamics of such driven-dissipative atom-cavity systems is well described by Eq.~(\ref{eq:photon_move_eq}) and the following Gross-Pitaevski (GP)-like equations for the macroscopic atomic wave functions
\begin{equation}
i\partial_t \Psi_1=H_{0,1}\Psi_1 \quad \text{and} \quad
i\partial_t \Psi_2=H_{0,2}\Psi_2.
\label{eq:GP_wave_function}
\end{equation}
The steady-state of $\Psi_j$ satisfies $i\partial_t \Psi_j=\mu_j\Psi_j$ with $\mu_j$ being the chemical potential for component $j$. To obtain the stationary states, we seek self-consistent solutions via the imaginary time propagation method \cite{Nagy2008}. $\alpha$ in Eq.~(\ref{eq:GP_wave_function}) is replaced with Eq.~(\ref{eq:stead_state_alpha}), and the integral in Eq.~(\ref{eq:stead_state_alpha}) is evaluated in every step of the time evolution. The normal phase and the superradiant phase can be discriminated by examining the cavity photon field order parameter $\alpha$, which vanishes in the normal phase while develops a finite value in the superradiant phase.

By substituting the expression for $\tilde{\Delta}_c=\Delta_c-\langle\sum_i\mathrm{sgn}(\Delta_{a,j})\int U_j(\mathbf{r})\hat{N}_j\mathrm{d}\mathbf{r}\rangle$ into Eq.~(\ref{eq:critical_effective_cavity_detuning}), we obtain the critical cavity detuning. The second term on the right-hand side of the expression for $\tilde{\Delta}_c$ is responsible for shifting the upper edge of the phase boundary.  We map out the phase diagram spanned by $V_0$ and $\Delta_c$ in Fig.~\hyperref[fig3]{3(b)}. The green line represents the phase boundary derived from the modified form of Eq.~(\ref{eq:critical_effective_cavity_detuning}) after substitution, while the numerical phase regions are obtained from the self-consistent approach described above. The mean-field prediction for the phase boundary is in good agreement with the numerical analysis. Notably, a minimum value of $|\Delta_c|$ exists at the top of the phase boundary. For $|\Delta_c|$ below this minimum, the system becomes unstable, as the ground-state energy of the system diverges with increasing order parameter $\alpha$.

%We map out the phase diagram spanned by $V_0$ and $\Delta_c$ in Fig.~3, in which the green line represents the phase boundary derived from Eq.~(\ref{eq:critical_effective_cavity_detuning}) and the numerical phase region is obtained by the self-consistent approach described above. Neglecting the minor horizontal shifts at the boundary caused by incomplete consideration of momentum occupation, we conclude that the analytical result for the phase boundary is in good agreement with the numerical result. We notice that the phase diagram we obtained resembles that of a single component BEC under red detuning \cite{Baumann2010}. There exists a minimum of $|\Delta_c|$, below which the system becomes unstable toward collapse. The phase boundary between the normal phase and the superradiant phase is linear. As $V_0$ increases, the threshold for phase transition shifts toward smaller values of $|\Delta_c|$: Once the system is in the superradiant phase, further increasing of $|V_0|$ will not drive the system out of it. In particular, we find that for a single component with blue detuning~\cite{Zupancic2019}, the superradiant region coincides with the unstable region of our system. This behavior indicates that the phase transition is mainly governed by the red detuning component.
% In addition, for the phase transition to occur, $V_0$ must exceed a minimum value that ensures $\mathcal{F}>\kappa/2$.

\begin{figure}[t]
    \includegraphics[width=1\linewidth]{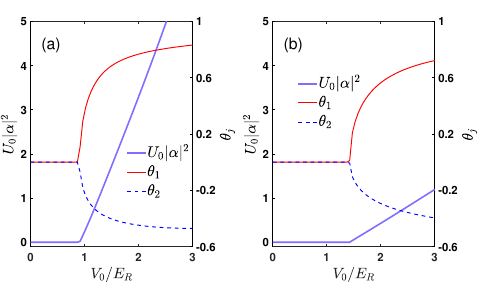}
    \caption{The evolution of the order parameters as a function of the tuning parameter $V_0$. we decompose the order parameter $\Theta$ into a set of components $\theta_j$, where $\Theta=\sum_j\theta_j$. (a) Here, we set $\Delta_c=2\pi\times-15 \mathrm{MHz}$ and (b) we set $\Delta_c=2\pi\times-24 \mathrm{MHz}$.}
    \label{fig4}
\end{figure}

\begin{figure}[t]
    \includegraphics[width=1\linewidth]{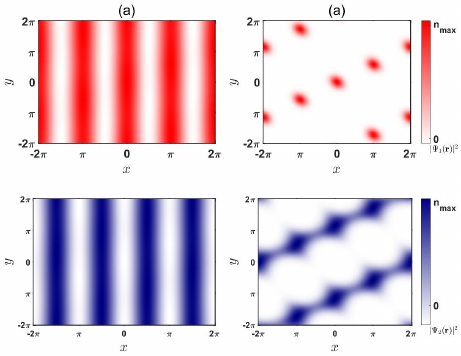}
    \caption{The density modulations of the two components are presented as follows: (a) the left column is in the normal phase, $V_0=1E_R$ and (b) the right column is in the supperradiant phase, $V_0=5E_R$. The cavity detuning  is fixed at $\Delta_c=2\pi \times-17 \mathrm{MHz}$. Red and blue colors represent density modulations under red-detuned and blue-detuned component conditions, respectively.}
    \label{fig5}
\end{figure}

The evolution of the order parameters with respect to the tuning parameter $V_0$ is shown in Fig.~\ref{fig4}. Following the same approach, we decompose the order parameter $\Theta$ into a set of components $\theta_j$, with the constraint $\Theta=\sum_j\theta_j$. Firstly, the finite value of $\alpha$ accompanied by finite values of $\theta_j$ indicates the simultaneous occurrence of atomic self-organized pattern formation and the superradiant phase transition in the optical cavity~\cite{Baumann2010}.
%The superradiant phase can also be referred to as the self-ordering phase \cite{Baumann2011}.
For $\Delta_c=2\pi\times-15 \mathrm{MHz}$, the transition point is located at $V_0/E_R=0.90$; For $\Delta_c=2\pi\times-24\mathrm{MHz}$, the transition point occurs at $V_0/E_R=1.41$. Specifically, the signs of $\theta_1 $ and $\theta_2$ are mutually opposite, indicating that  the density maxima of the two components do not coincide~\cite{Landini2018}. Let us turn to Fig.~\ref{fig5}, which shows a schematic diagram of normalized density in the normal and superradiant phases. The global structures of the two condensates are always phase separated in real space \cite{Ali2022}: the two components exhibit the alternate stripe pattern in the normal phase, while they form Bragg gratings with the different symmetry centers in the superradiant phase.

\section{Roton-Type Mode Softening}
Finally, let us analyze the excitation spectrum across phase transition between normal and superradiant phases. Considering the deviations from the steady state $\hat{a}(t)=\alpha+\delta\alpha(t)$ and $\Psi_j(\mathbf{r},t)=\big[ \Psi_{j}(\mathbf{r})+\delta\psi_j(\mathbf{r},t) \big]e^{-i\mu_jt/\hbar}$, then substituting them into Eqs.~(\ref{eq:photon_move_eq}) and (\ref{eq:GP_wave_function}) and linearizing the equations in $\delta\alpha$ and $\delta \psi_j$, one gets
\begin{equation}
\begin{aligned}
i\hbar\delta\dot{\alpha}=&A\delta\alpha+ \sum_j\mathrm{sgn}(\Delta_{a,j})N_j\Bigg[ \alpha\big( \langle \Psi_{j}|U_j(\mathbf{r})|\delta\psi_j\rangle+H.c. \big)\\
& +\big( \langle \Psi_{j}|\eta_j(\mathbf{r})|\delta\psi_j\rangle+H.c. \big) \Bigg],
\label{eq:fluctuate_alpha}
\end{aligned}
\end{equation}
\begin{equation}
\begin{aligned}
i\hbar\delta\dot{\psi}_j=&H_j\delta\psi_j+\mathrm{sgn}(\Delta_{a,j})\Big[ U_j(\mathbf{r})\big( \alpha^*\delta\alpha+\alpha\delta\alpha^* \big)\\
&+\eta_j(\mathbf{r})\big( \delta\alpha^*+\delta\alpha \big) \Big]\Psi_{j},
\label{eq:fluctuate_particle}
\end{aligned}
\end{equation}
where we have defined $A=-\tilde{\Delta}_c-i\kappa$ and $H_j=H_{0,j}-\mu_j$. As linearized Eqs.~(\ref{eq:fluctuate_alpha}) and (\ref{eq:fluctuate_particle}) couples $\delta\alpha$ and $\delta\psi_j$ to their complex conjugates, we search the solution in the form $\delta\alpha(t)=e^{-i\omega t/\hbar}\delta\alpha_{+}+e^{-i\omega^* t/\hbar}\delta\alpha^*_{-}$ and
$\delta\psi_{j}(\mathbf{r},t)=e^{-i\omega t/\hbar}\delta\psi_{j,+}(\mathbf{r})+e^{i\omega^* t/\hbar}\delta\psi^*_{j,-}(\mathbf{r})$. A set of coupled Bogoliubov-type equations are derived by substituting them into Eqs.~(\ref{eq:fluctuate_alpha}) and (\ref{eq:fluctuate_particle}) and explicitly writing the equations for both the positive- and negative-frequency components of the quantum fluctuations:
\begin{equation}
\begin{aligned}
\omega\delta\alpha_{+}=&\sum_j\Big\{\mathrm{sgn}(\Delta_{a,j})N_j\big[\langle\delta\psi_{j,-}^*|U_j(\mathbf{r})\alpha+\eta_j(\mathbf{r})|\Psi_{j}\rangle\\
&+\langle\Psi_{j}|U_j(\mathbf{r})\alpha+\eta_j(\mathbf{r}))|\delta\psi_{j,+}\rangle\big]\Big\}+A\delta\alpha_{+},\\
-\omega^*\delta\alpha_{-}^*=&\sum_j\Big\{\mathrm{sgn}(\Delta_{a,j})N_j\big[\langle\delta\psi_{j,+}|U_j(\mathbf{r})\alpha+\eta_j(\mathbf{r})|\Psi_{j}\rangle\\
&+\langle\Psi_{j}|U_j(\mathbf{r})\alpha+\eta_j(\mathbf{r}))|\delta\psi_{1,-}^*\rangle\big]\Big\}+A\delta\alpha_{-}^*,\\
\omega\delta\psi_{j,+}=&H_j\delta\psi_{j,+}+\mathrm{sgn}(\Delta_{a,j})\big[U_j(\mathbf{r})\alpha^*+\eta_j(\mathbf{r})\big]\Psi_{j}\delta\alpha_{+}\\
&+\mathrm{sgn}(\Delta_{a,j})\big[U_j(\mathbf{r})\alpha+\eta_j(\mathbf{r}))\big]\Psi_{j}\delta\alpha_{-},\\
-\omega^*\delta\psi_{j,-}^*=&H_j\delta\psi_{j,-}^*+\mathrm{sgn}(\Delta_{a,j})\big[U_j(\mathbf{r})\alpha+\eta_j(\mathbf{r})\big]\Psi_{j}\delta\alpha_{+}^*\\
&+\mathrm{sgn}(\Delta_{a,j})\big[U_j(\mathbf{r})\alpha^*+\eta_j(\mathbf{r}))\big]\Psi_{j}\delta\alpha_{-}^*.
\end{aligned}
\end{equation}
This set of equations can be written in a matrix form $\omega\mathrm{f}=\mathcal{M}\mathrm{f}$ with $\mathrm{f}=(\delta\alpha_{+},\delta\alpha_{-},\delta\psi_{1,+},\delta\psi_{1,-},\delta\psi_{2,+},\delta\psi_{2,-})^{T}$. The excitation energy corresponds to the solution of the eigenvalue problem for the Bogoliubov matrix $\mathcal{M}$.

\begin{figure}[t]
    \includegraphics[width=1\linewidth]{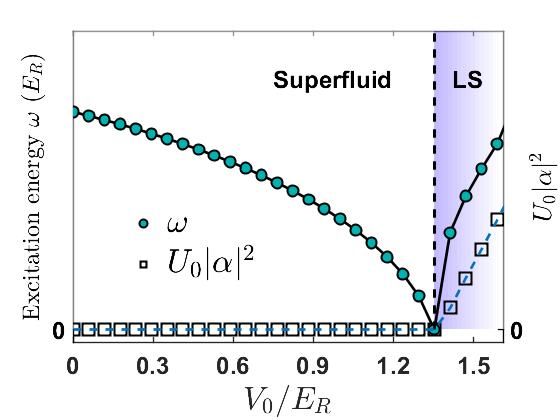}
    \caption{Numerical results for the excitation energy $\omega$ and  the intracavity depth $U_0|\alpha|^2$ as functions of the tuning parameter $V_0$ are presented. A roton-type mode softening occurs when the  $U_0|\alpha|^2$ becomes non-zero. This behavior signals a phase transition from the superfluid to lattice supersolid phase. Here we set $\Delta_c=2\pi \times-23\mathrm{MHz}$, and we find the critical point $V_0^{\mathrm{crit}}\approx1.353$ for phase transition from normal to superradiant phase.  $\textrm{LS}$ represents the lattice supersolid phase.}
%Numerical results of excitation energy $\omega$ as a function of the tuning parameter $V_0$. It possesses a roton-type softened mode as $V_0$ increases. Here we set $\Delta_c=2\pi \times-23\mathrm{MHz}$.}
    \label{fig6}
\end{figure}

Fig. \ref{fig6} shows the excitation energy $\omega$ and the intracavity depth $U_0|\alpha|^2$. A rotor-type mode softening is observed at the critical point where $U_0|\alpha|^2$ becomes finite. This softening signals a phase transition from normal superfluid to lattice supersolid phase, which is driven by enhancing the light-atom interaction through the parameter $V_0$~\cite{Pomeau1994,Mottl2012}. The softened roton is the Goldstone mode of the broken two discrete $\mathcal{Z}_2$ symmetries. Here, the atomic field operator $\hat{\Psi}_j$ can be expanded  using plane waves as
\begin{equation}
\Psi_j=\langle\mathbf{r}|\Psi_j^1(0)\rangle\hat{b}_{j,0}^{(1)}+\sum_{\mathbf{k},n}\langle\mathbf{r}|\Psi_j^n(\mathbf{k})\rangle \hat{b}_{j,\mathbf{k}}^{(n)}
\label{eq:expanded_using_plane_waves}
\end{equation}
where $\hat{b}_{j,\mathbf{k}}^{(n)}$ is the  bosonic annihilation operators for states $|\Psi_j^n(\mathbf{k})\rangle$. The particle number conservation $\hat{b}^{{(1)} \dagger}_{j,0}\hat{b}_{j,0}^{(1)}+\sum_{\mathbf{k},n}\hat{b}_{j,\mathbf{k}}^{{(n)} \dagger}\hat{b}_{j,\mathbf{k}}^{(n)}=N_j$ should  be satisfied. Substitution of Eq.~(\ref{eq:expanded_using_plane_waves}) into Eq.~(\ref{eq:effective_Hamiltonian_in_a_frame_rotating}) results in the effective Hamiltonian taking the form in the momentum space
\begin{equation}
\begin{aligned}
\hat{H}_{\mathrm{eff}}=&-\hbar\tilde{\Delta}_c\hat{a}^\dagger\hat{a}+\sum_{j=1,2}\sum_{\mathbf{k},n}\Bigg[\Big(E^n_{j}(\mathbf{k})-E^{1}_{j}(0)\Big)\hat{b}_{j,\mathbf{k}}^{(n) \dagger}\hat{b}_{j,\mathbf{k}}^{(n)}\\
&+(\hat{a}^\dagger+\hat{a})\Big(\nu^{(n)}_{j,k}\hat{b}_{j,\mathbf{k}}^{{(n)} \dagger}\hat{b}_{j,0}^{(1)}+H.c.\Big)\Bigg]
\label{eq:the_Dicke_Hamitonian}
\end{aligned}
\end{equation}
where $\nu^{(n)}_{j,\mathbf{k}}=\langle\Psi_j^{n}(\mathbf{k})|\mathrm{sgn}(\Delta_{a,j})\eta_j(\mathbf{r})\hat{N}_j|\Psi_j^{1}(\mathbf{0}) \rangle$. After that, Eq.~(\ref{eq:energy_expressed_Theta}) can always be obtained by implementing the mean-field theory and seeking a steady state for atomic fields~\cite{QinWei2024}.  Obviously, the Eq.~(\ref{eq:the_Dicke_Hamitonian}) exactly possesses a “composite” $\mathcal{Z}_2$ symmetry, as it is invariant under the simultaneous transformation of $\hat{a}\to-\hat{a}$ and $\hat{b}_{j,\mathbf{k}}^{(n)}\to-\hat{b}_{j,\mathbf{k}}^{(n)}$. The system spontaneously breaks this $\mathcal{Z}_2$ symmetry during the transition from the normal phase to superradiant phase~\cite{leonard_thesis_2017}.
%The excitation energy $\omega$ is presented in Fig. \ref{fig6}, showing a softened roton-type mode across a continuum of finite $V_0$. As $V_0$ approaches its critical value, the excitation energy tends toward zero, marking the phase transition between the normal phase and the superradiant phase. Especially, the tendency of the system to crystallize can be predicted by the presence of the roton mode in the excitation energy $\omega$~\cite{Pomeau1994}, while the roton-type mode softening phenomenon itself signifies the superfluid-to-supersolid transition \cite{Mottl2012}.
\section{Summary and Outlook}
In summary, we have demonstrated the superradiant phase transition in a two-component BEC with distinct atomic detunings, confined in an optical cavity and driven by a transverse pump laser. The susceptibility as a function of pump strength $V_0$ exhibits monotonicity, primarily influenced by the red-detuned component. Further investigation reveals that the superradiant properties of the system\textemdash including the phase boundary and the Bogoliubov excitation spectrum\textemdash closely resemble those of the single-component red-detuned case. The phase diagram features a minimum in $|\Delta_c|$, below which the system enters an unstable region due to the divergence of the ground-state energy with increasing order parameter $\alpha$.
%However, the system becomes unstable when approaching the transition parameters of a single-component blue-detuned BEC.
Due to the distinct attractive or repulsive potentials induced by red or blue atomic detuning, we find that there exists a natural phase separation in real space between the two components.

A key requirement for the experimental feasibility is that two bosonic species of the same mass form Bose-Einstein condensates with distinct pump-atom detunings $\Delta_{a,j}$. The combination of $^{87}\mathrm{Rb}$ and $^{88}\mathrm{Sr}$ is a suitable candidate~\cite{LuoChengyi2024}, due to their large difference in transition frequencies. For $^{87}\mathrm{Rb}$, we consider the D2 line from $5\mathrm{S}_{1/2}$ to $5\mathrm{P}_{3/2}$ at a wavelength of 780 nm~\cite{Torre2013}. For $^{88}\mathrm{Sr}$, we utilize the narrow-linewidth transition from $^1\mathrm{S}_0$ to $^3\mathrm{P}_1$ at 689 nm~\cite{Song2025}. The substantial transition energy difference of approximately 50.6 THz allows for the realization of a binary BEC with red and blue detunings by tuning the pump frequency $\omega_p$ in the far-detuned regime where $|\Delta_{a,j}|$ is much larger than their transition frequencies. We emphasize that the assumption of equal masses and atom numbers for the two BEC components is not essential to this study, whose primary goal is to explore what kind of superradiant phase diagrams and superradiant characteristics are produced by a binary condensate with distinct atomic detunings. Generalizations are straightforward: if the masses $m_j$ differ, the recoil energy $E_R=\hbar^2k^2_0/2m_j$ can be redefined by using one of two masses; if the atomic numbers $N_j$ differ, the component susceptibilities $f_j$ are directly affected (Eq.~(\ref{eq:susceptibility_function}) contains the atomic number operators $\hat{N}_j$). Furthermore, the relative angle between the cavity and pump fields influences the emergent density patterns in the superradiant phase by altering the coupling of atoms to different momentum states~\cite{ZUPANCIC_thesis_2020}. At a relative angle of $90^\circ$, the atoms of the two components localize separately on the even and odd sublattices of the resulting checkerboard potential~\cite{Baumann2010}.

Our model can be generalized to short-range interacting systems to investigate the two-component extended Bose-Hubbard model \cite{Carl2023}. The implementation of asymmetric pump lasers enables theoretical studies of PT-symmetry breaking \cite{Baur2025}. Furthermore, an interesting extension of this work would involve mapping out finite-temperature phase diagram \cite{ZhangYuanWei2013}. Our work advances the theoretical understanding of superradiant phase transitions in such systems, and our predictions provide feasible experimental verification. The cavity field can be measured with a balanced heterodyne setup \cite{Baumann2011}. From an applied perspective, our model shows potential applications for realizing controllable optical switches leveraging the phase transition between normal to superradiance.

\section*{Acknowledgments}
This work is supported by the National Natural Science Foundation of China under Grants No.~12174055 and No.~11674058, and by the Natural Science Foundation of Fujian Province under Grant No.~2025J01658.
%\bibliography{RefLin}

\end{document}